\documentclass[journal=jacsat,manuscript=article]{achemso}

\usepackage[utf8]{inputenc}
\usepackage{graphicx}
\usepackage{dcolumn}
\usepackage{multirow}
\usepackage{booktabs}
\usepackage{float}
\usepackage{bm}

\usepackage{siunitx}
\DeclareSIUnit{\rydberg}{Ry}
\DeclareSIUnit{\electronmass}{\ensuremath{m_0}}
\usepackage[version=4]{mhchem}
\usepackage[labelfont=bf, font=small, labelsep=period]{caption}
\usepackage{hyperref}
\hypersetup{
    colorlinks=true,
    linkcolor=blue,
    filecolor=magenta,      
    urlcolor=blue,
    citecolor=blue,
    allcolors = blue
    }
 \usepackage{physics}   
 \usepackage{indentfirst}

\author{Enesio Marinho Jr.}
 \email{enesio.marinho@unesp.br}
\affiliation{Department of Physics and Chemistry, School of Engineering, São Paulo State University (UNESP),
Ilha Solteira, São Paulo 15385-007, Brazil}

\author{Alexandre C. Dias}
 \affiliation{Institute of Physics and International Center of Physics, University of Brasília, Brasília, Distrito
Federal 70919-970, Brazil}
\author{Luiz A. Ribeiro Jr.}
 \affiliation{Institute of Physics, University of Brasília, Brasília, Distrito Federal 70919-970, Brazil}
\affiliation{Computational Materials Laboratory, LCCMat, Institute of Physics, University of Brasília, Brasília, Distrito Federal 70919-970, Brazil}
\author{Maurizia Palummo}
\affiliation{INFN \& Department of Physics, University of Rome Tor Vergata, Via della Ricerca Scientifica 1, 
00133 Rome, Italy}
 \author{Cesar E.~P.~Villegas}
 \email{cesar.perez@upn.edu.pe}
 \affiliation{Departmento de Ciencias, Universidad Privada del Norte, 15434 Lima, Peru}

\title{Exciton dynamics and high-temperature excitonic superfluidity in S-doped graphyne}

\begin{document}

\begin{tocentry}
    \includegraphics[width=\linewidth]{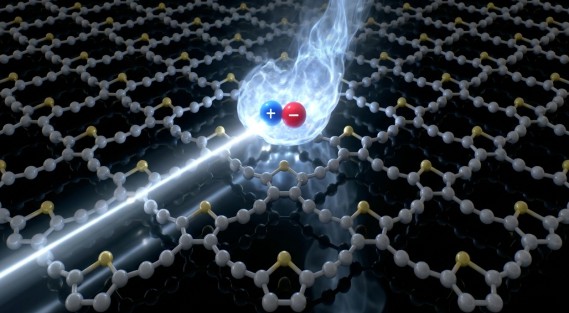}
\end{tocentry}

\begin{abstract}
 S-doped graphyne (S-GY) is a recently synthesized two-dimensional graphyne-based carbon allotrope that provides a promising platform for exciton engineering and coherent many-body phases. Here, we investigate the quasiparticle electronic structure, optical response, and exciton dynamics of monolayer S-GY using the G$_0$W$_0$ approximation and the Bethe--Salpeter equation (BSE). Quasiparticle corrections increase the fundamental band gap from $\SI{0.88}{\electronvolt}$ (PBE) to $\SI{1.95}{\electronvolt}$, while slightly reducing the carrier effective masses. The BSE optical response reveals strongly bound excitons, with the lowest bright exciton exhibiting a binding energy of $\SI{0.72}{\electronvolt}$, as well as a nearly degenerate dark exciton within the thermal energy scale. Analysis of exciton wavefunctions in reciprocal space confirms a hydrogenic Rydberg series with well-defined angular-momentum character, and radiative lifetimes in the nanosecond range at room temperature, comparable to those in transition-metal dichalcogenide monolayers. Finally, we construct the excitonic phase diagram and estimate a crossover density of ${\sim}6\times10^{12}~\text{cm}^{-2}$, below which the exciton gas behaves as a dilute Bose system, and the Berezinskii--Kosterlitz--Thouless (BKT) superfluid phase becomes accessible. We estimate a maximum BKT transition temperature of ${\sim}\SI{143}{\kelvin}$ in the freestanding limit for the 1s exciton, indicating that monolayer S-GY may provide favorable conditions for high-temperature excitonic superfluidity in graphyne-based materials.
\end{abstract}

\section{Introduction}
Graphyne-based materials (GBMs) have emerged as a prominent class of layered carbon allotropes, exhibiting bonding motifs and physical properties distinct from those of conventional $sp^2$ carbon networks \cite{Kang-ACSAMI-2019, Chandran_JMCA-2026}. First proposed by Baughman \textit{et al.} \cite{baughman-JCP-1987}, graphyne (GY) consists of coexisting $sp$- and $sp^2$-hybridized carbon atoms arranged in periodic two-dimensional frameworks interconnected by acetylenic linkages. In contrast to graphene, this mixed hybridization introduces intrinsic porosity and enhanced structural tunability. Distinct graphyne polymorphs emerge from differences in the number of carbon atoms forming the smallest and next-smallest rings connected by acetylenic bridges, giving rise to $\alpha$, $\beta$, $\gamma$, and $6,6,12$-graphyne, among otherpolymorphs \cite{baughman-JCP-1987}. Of these allotropes, only monolayer $\gamma$-graphyne is a direct band-gap semiconductor, whereas the remaining phases exhibit Dirac semimetal behavior \cite{Li_CSR-2014}.

Structural diversification beyond the benzenoid GBMs has further expanded the graphyne family. A pentagon-based graphyne (PBG), termed cp-graphyne, was theoretically proposed in Ref.~\cite{Nulakani-ACSO-2017}, where six-membered carbon rings are replaced by five-membered ones, leading to a semimetallic electronic structure characterized by double distorted Dirac points. Subsequently, imidazole-graphyne was computationally designed as a semiconducting PBG constructed from experimentally synthesized five-membered imidazole units connected by acetylenic linkers, exhibiting a direct band gap of approximately $\SI{1.10}{\electronvolt}$ \cite{Zhou-PCCP-2021}. More recently, the first PBG derivative was experimentally realized through a bottom-up, solvent-free mechanochemical reaction between calcium carbide and tetrabromothiophene, yielding S-doped graphyne (S-GY) \cite{Li-SPT-2025}. This material has demonstrated promising performance in Hg(II) adsorption and electrochemical energy storage. In our recent first-principles study \cite{Aparicio-Huacarpuma_PRB-2025}, we showed that monolayer S-GY is thermodynamically stable and exhibits pronounced elastomechanical anisotropy. At the hybrid-functional DFT level, it presents an indirect band gap of approximately $\SI{1.3}{\electronvolt}$, broken electron–hole symmetry, and van Hove singularities located away from the Fermi level.

The reduced dimensionality and weak dielectric screening inherent to GBMs strongly enhance many-body interactions, making their optical response highly sensitive to quasiparticle and excitonic effects. Graphdiyne provides a representative example \cite{Luo_PRB-2011}, for which combined experimental measurements and GW plus Bethe–Salpeter equation (BSE) calculations reveal substantial quasiparticle band-gap renormalization and pronounced excitonic resonances, with binding energies exceeding $\SI{0.55}{\electronvolt}$ and exhibiting both Wannier–Mott and Frenkel characteristics. Similarly, theoretical investigations of graphyne and related two-dimensional alkynyl networks predict sizable exciton binding energies and strong optical absorption features \cite{Huang_PRB-2013}. Quasiparticle calculations for graphyne yield a band gap of about 1.4 eV and an exciton binding energy near 400 meV, accompanied by significant singlet–triplet splitting and near-infrared absorbance exceeding $\SI{6}{\percent}$. To rationalize these excitonic spectra, a modified one-parameter hydrogenic model has been proposed, incorporating anisotropic confinement and electron–hole exchange effects.

These pronounced many-body effects naturally motivate the exploration of collective excitonic phenomena. Quantum phases of excitons have long been investigated as a fundamental manifestation of electron-hole correlations \cite{Blatt_PR-1962, Keldysh-Kozlov_ZETF-1968}. As a remarkable experimental result, Morita \textit{et al.}~\cite{Morita_NC-2022} reported the observation of macroscopic quantum coherence of 1s paraexcitons in bulk \ce{Cu2O} below ${\sim}\SI{400}{\milli\kelvin}$, directly visualizing the exciton condensate via mid-infrared induced absorption imaging and revealing a condensate fraction of order $\SI{1.6e-2}{}$ that is undetectable by conventional luminescence spectroscopy. Because excitons possess relatively small effective masses, often comparable to the free electron mass, they can, in principle, reach quantum degeneracy at temperatures substantially higher than those achievable in atomic gases \cite{Butov_Nature-2002}.  The critical temperature depends not only on the exciton mass but also on the binding energy, since strongly bound excitons are more robust against thermal dissociation. In this regard, atomically thin two-dimensional materials are particularly attractive platforms, as the quantum confinement enhances Coulomb interactions, yielding exciton binding energies significantly larger than those in conventional bulk semiconductors \cite{Wang_RMP-2018}. A primary obstacle to realizing collective excitonic phases is the finite lifetime of excitons, as the quasiparticle lifetime must be longer than the thermalization time to enable the formation of a quasi-equilibrium population and, consequently, macroscopic phase coherence. \cite{Snoke_Science-2002}. This issue can be addressed by selecting or designing systems with intrinsically long exciton lifetimes, such as van der Waals heterostructures \cite{Rivera_NC-2015}, that host interlayer excitons, or materials containing dark excitons \cite{Combescot_PRL-2007}, given that dark states are optically inactive due to selection rules governing dipole transitions, and their lifetimes are significantly larger than those of the bright excitons, enabling the buildup of the high exciton densities required for coherent phases. 

Macroscopic excitonic quantum phases in two-dimensional carbon-based materials have attracted sustained interest. It has been experimentally realized in graphene bilayers \cite{Perali_PRL-2013,Li_NP-2017,Liu_Science-2022}. Despite this remarkable achievement, the relatively weak exciton binding energy in graphene limits the characteristic transition temperature to only a few kelvin. The prospect of achieving more robust condensates in semiconducting derivatives was subsequently examined by Cudazzo et al. \cite{Cudazzo_PRL-2010}, who demonstrated that stronger binding significantly enlarges the stability window. More recently, first-principles GW–BSE calculations have predicted the possibility of high-temperature excitonic coherent phases in holey-graphyne \cite{Yue_PRB-2025}, a newly synthesized GBM \cite{Liu_Matter-2022}.

In this work, we employ many-body perturbation theory within the GW–BSE framework to provide a comprehensive and predictive description of quasiparticle band-gap renormalization and excitonic effects in the optical response of S-GY. Beyond establishing its excited-state electronic structure, we quantitatively assess the radiative lifetimes of the excitonic states and delineate the physical conditions required for the emergence of correlated electron–hole quantum phases, including excitonic superfluidity. Our results indicate that monolayer S-GY could sustain an excitonic superfluid phase at temperatures as high as $\SI{143}{\kelvin}$, placing it among the most promising two-dimensional platforms for the realization of elevated-temperature excitonic quantum coherence.

\section{\label{sec:methodology}Computational Details}

Our ground-state calculations are based on the density-functional theory (DFT) \cite{Hohenberg1964,Kohn1965} as implemented in the \textsc{quantum espresso} package \cite{Kresse-PRB-1993, Kresse-PRB-1996}. For the exchange–correlation potential, we employ the Perdew–Burke–Ernzerhof (PBE) functional within the generalized gradient approximation (GGA) \cite{PBE-PRL-1996}, with norm-conserving \textsc{PseudoDojo} pseudopotentials \cite{vanSetten_CPC-2018} to model the electron-ion interactions. We expand the valence electron wavefunctions in plane-wave basis sets with a kinetic energy cutoff of \SI{90}{\rydberg}.  The structural full-relaxation and self-consistent calculations are carried out with a Monkhorst–Pack $\vb{k}$-point grid of $12\times 12\times 1$. A vacuum spacing of $\SI{15}{\angstrom}$ was introduced along the out-of-plane direction to avoid spurious interactions due to periodic boundary conditions.

The well-known underestimation of the band gap within density functional theory is overcome by employing the GW formalism~\cite{hedin-PR-1965,Hydertsen-Louie-PRB-1986}, while optical response and excitonic properties are obtained by solving the Bethe--Salpeter equation (BSE)~\cite{BSE_PR-1951,Blase_JPCL-2020}. All those excited-state calculations within the many-body perturbation theory were performed with the \textsc{yambo} code~\cite{Marini_CPC-2009,Sangalli_JPCM-2019}. Quasiparticle corrections to the DFT eigenvalues are evaluated within the single-shot GW scheme (G$_0$W$_0$)~\cite{Aryasetiawan_RPP-1998}. The G$_0$W$_0$ quasiparticle energies were calculated with 300 bands to construct the screened Coulomb potential and 200 bands to expand the Green’s function, within the plasmon-pole approximation for the dynamical dielectric response function \cite{Rojas_PRL-1995}, and a $\vb{G}$-vector energy cutoff of $\SI{8}{\rydberg}$. A kinetic energy cutoff of $\SI{12}{\rydberg}$ was used to evaluate the exchange part of the self-energy. To speed up the convergence of GW quasiparticle band gaps with respect to the $\vb{q}$ grid, we employ the W-av method \cite{Guandalini_nCM-2023}. Including $\SI{1000000}{}$ random $\vb{q}$-points to perform the Monte Carlo integration, with a cutoff of $\SI{1.5}{\rydberg}$ for the Coulomb interaction and $\SI{2.0}{\rydberg}$ to numerically integrate the screening, we obtain sufficient accurate quasiparticle band gaps for a $\vb{q}$-point grid of $12\times12\times1$. A terminator scheme is used to speed up the convergence of the electronic Green’s function with respect to the empty states \cite{Bruneval_Gonze_PRB-2008}. Based on our systematic convergence tests, using the specified GW features allows computing quasiparticle band gaps of monolayer S-GY with a total error of less than $\SI{50}{\milli\electronvolt}$. 

The optical absorption, including excitonic effects, is computed by solving the BSE on top of G$_0$W$_0$ results, within the Tamm--Dancoff approximation~\cite{Striani_RNC-1988}. A Coulomb interaction truncated in a slab geometry is applied to suppress unwanted interactions between periodic replicas.  We use $80$ bands and a $\vb{G}$-vector cutoff of $\SI{8}{\rydberg}$ for the static dielectric screening. We include the six highest-occupied valence
bands and the six lowest-unoccupied conduction bands to diagonalize the BSE Hamiltonian, along with a $\Gamma$-centered $\vb{k}$-point grid of $24\times24\times1$. To construct the statically screened BSE kernel, the electron-hole exchange part was computed with a $\vb{G}$-vector cutoff of $\SI{12}{\rydberg}$, while the electron-hole
attraction part was calculated using a block size of $\SI{6}{\rydberg}$. By setting the BSE parameters to the specified values, the exciton binding energy of the most intense resonance in the optical spectrum is expected to converge within $\SI{50}{\milli\electronvolt}$. At the independent quasiparticle level (IQP), the dielectric matrix is computed within the random-phase approximation (RPA)~\cite{Bohm_PR-1953}.

\section{\label{{sec:results}}Results and Discussion}
\subsection{GW Quasiparticle band structure}
\begin{figure*}[t!]
    \centering
    \includegraphics[width=\linewidth]{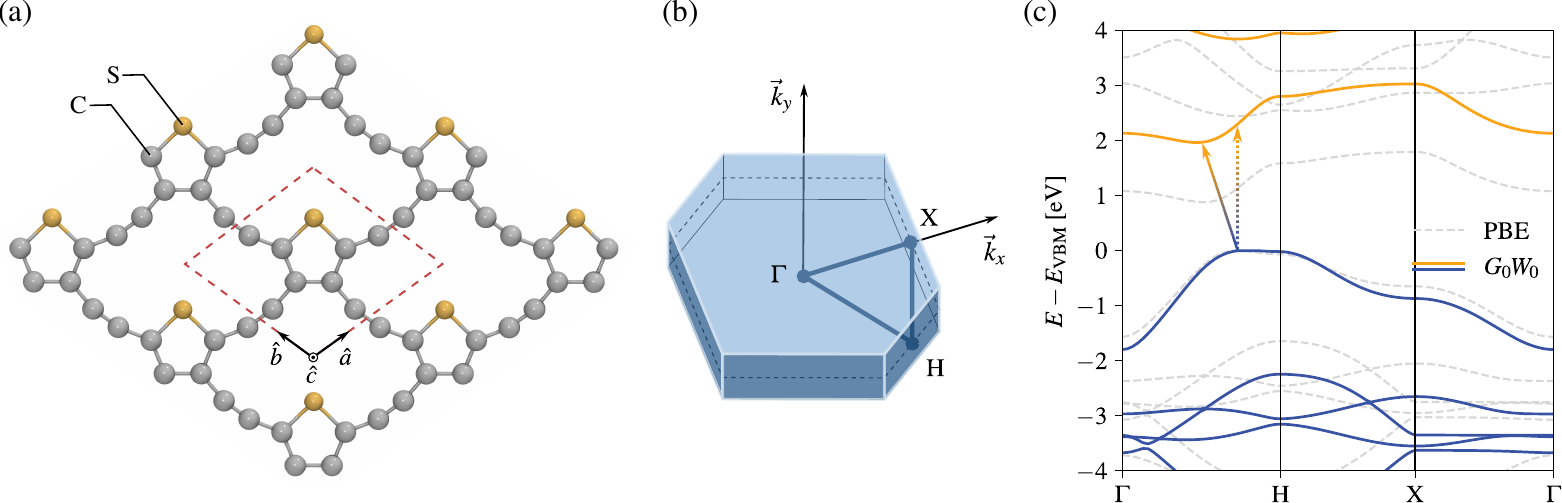}
    \caption{Real- and reciprocal-space representations, and electronic properties of monolayer S-doped graphyne (SGY). (a) Crystal structure, with the unit cell indicated by the parallelogram. (b) Corresponding Brillouin zone with high-symmetry $\vb{k}$ points. (c) Electronic band structure of monolayer SGY, where solid lines correspond to GW bands and dashed lines to DFT-PBE results. In both cases, the respective VBM energy is set to zero.}
    \label{fig:structure-BZ-bands}
\end{figure*}
The fully-relaxed crystalline structure of monolayer S-GY is shown in Fig.~\hyperref[fig:structure-BZ-bands]{\ref*{fig:structure-BZ-bands}(a)}. In contrast to graphene and $\gamma$-graphyne, which belong to the high-symmetry hexagonal wallpaper group $p6m$, the S-GY monolayer exhibits a markedly reduced symmetry. Graphene and $\gamma$-graphyne possess a hexagonal Bravais lattice with sixfold rotational symmetry ($C_6$), multiple vertical mirror planes, and inversion symmetry, which enforce high degeneracy and in-plane isotropy in the electronic structure. On the other hand, the fully relaxed single-layer S-GY crystallizes in the wallpaper group \textit{pm}, with $a=b=\SI{6.30}{\angstrom}$ and $\gamma \approx \SI{110}{\degree}$. The structure is characterized by a rhombic lattice and a single mirror plane ($m$) passing through the sulfur atoms. As a consequence, the only non-trivial point-symmetry operation is this mirror reflection \cite{symmetry_comment}. Notably, the substitution of carbon by sulfur, together with the presence of five-membered rings instead of conventional six-membered carbon rings, breaks all rotational and inversion symmetries. 

We proceed to analyze the quasiparticle band structure of the S-GY monolayer, computed within the G$_0$W$_0$ approximation on top of DFT-PBE. Fig.~\hyperref[fig:structure-BZ-bands]{\ref*{fig:structure-BZ-bands}(c)} shows the resulting quasiparticle band structure along the high-symmetry paths of the Brillouin zone (BZ) indicated in Fig.~\hyperref[fig:structure-BZ-bands]{\ref*{fig:structure-BZ-bands}(b)}.  The lowest direct and indirect band gaps are indicated by dashed and solid arrows in the GW band structure, respectively. At the DFT-PBE level, the S-GY monolayer exhibits an indirect band gap of $\SI{0.88}{\electronvolt}$, while the lowest direct gap is $\SI{1.13}{\electronvolt}$. The inclusion of quasiparticle corrections increases the fundamental gap to $\SI{1.95}{\electronvolt}$, with a corresponding direct gap of $\SI{2.19}{\electronvolt}$. This yields a quasiparticle band-gap renormalization of approximately $\SI{1.07}{\electronvolt}$ relative to PBE, which is typical for other 2D semiconductors \cite{Qiu_PRL-2013}, including the pristine GY monolayer \cite{Huang_PRB-2013}, and leads to quasiparticle corrections that improve agreement with experimental optoelectronic properties \cite{Ugeda_NM-2014}.

The effective masses are also modified by the inclusion of quasiparticle effects. At the PBE level, the hole effective mass at the valence-band maximum is 
$\SI{0.40}{\electronmass}$, while the electron effective mass at the conduction-band minimum is  $\SI{0.32}{\electronmass}$. When GW quasiparticle corrections are taken into account, these values decrease to $\SI{0.34}{\electronmass}$ for holes and $\SI{0.27}{\electronmass}$ for electrons. This reduction indicates a slight enhancement of the band dispersion arising from many-body effects, resulting in lighter carriers compared to the PBE description. These results, along with the reduced mass of the electron-hole pair and the exciton effective mass, are summarized in Table \ref{tab:bandgaps}.

\begin{table}[t]
\caption{Band-gap energies (in eV), electron and hole effective masses, together with the reduced mass of the electron-hole pair and the exciton effective mass (in units of the electron rest mass $m_0$), obtained at the DFT and GW levels for monolayer SGY.}
\label{tab:bandgaps}

\begin{tabular}{lcc}
\hline\hline
        & DFT-PBE & GW \\
\hline
$E^\text{i}_\text{g}$   & $0.88$    & $1.95$    \\
$E^\text{d}_\text{g}$  & $1.13$    & $2.19$    \\
$m^*_e$             & $0.32$    & $0.27$    \\
$m^*_h$              & $0.40$    & $0.34$    \\
$\mu_{eh}$             & $0.18$    & $0.15$    \\
$M_\text{S}$             & $0.72$    & $0.61$    \\\hline\hline
\end{tabular}
\end{table}

\begin{figure}[t!]
    \centering
    \includegraphics[width=.6\linewidth]{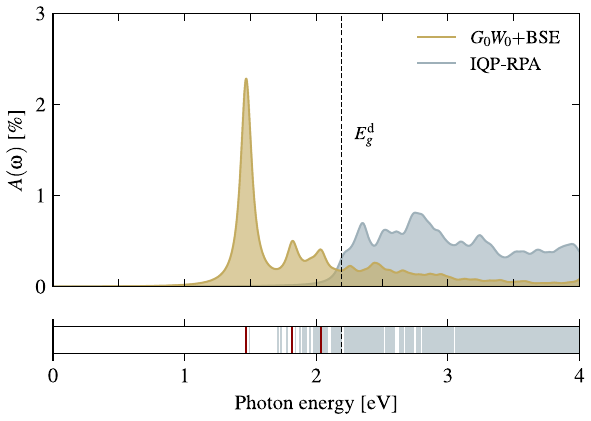}
    \caption{Absorption spectra for monolayer SGY obtained with the IQP-RPA (shaded area in blue) and BSE (shaded area in red). The GW fundamental band gap is marked with a vertical dashed line. In the inset below, the orange vertical lines indicate the bright excitons, while the grey lines depict the dark excitons.}
    \label{fig:optical-spectra}
\end{figure}

The optical absorption of S-GY is expressed in terms of the absorbance \cite{Yang_PRL-2009}
\begin{equation}
    A(\omega) = \frac{4\pi\omega}{c}\alpha_2(\omega)\,,
\end{equation}
where $\alpha_2(\omega)$ denotes the imaginary part of the polarizability per unit area. 
The resulting spectra are presented in 
Fig.~\hyperref[fig:optical-spectra]{\ref*{fig:optical-spectra}}.
The G$_0$W$_0$+BSE optical response, which includes excitonic effects, is compared 
to the spectrum where electron–hole interactions are neglected, 
calculated within the independent-particle random-phase approximation (IQP-RPA). Both optical spectra were computed for $\vb{E} \perp \hat{c}$ polarization of the incident light, where $\hat{c}$ denotes the out-of-plane direction. In our recent work \cite{Aparicio-Huacarpuma_PRB-2025}, we discuss the in-plane anisotropy of the BSE optical absorption spectrum, showing that this effect is prominent only at high energies, above the continuum limit. Once the optical absorption spectra have been computed considering only direct optical transitions (i.e., with $\vb{q} = \vb{0}$), the GW direct band-gap energy is indicated by a vertical dashed line, representing the continuum limit. Moreover, the exciton spectrum, together with the corresponding exciton oscillator strengths, is shown in the insets below. As a threshold to distinguish between bright and dark excitons, we define excitons with oscillator strengths smaller than $\SI{5}{\percent}$ of the most intense peak 
as dark excitons, whereas the others are classified as bright excitons. 

\subsection{Optical response and exciton spectrum}

As expected, accounting for electron–hole interactions gives rise to excitonic resonances in the absorption spectrum, yielding peaks that are both significantly redshifted and considerably more intense than their IQP-RPA counterparts. As a result, we identify three bound excitons associated with the three most prominent resonances in the BSE optical spectrum. The first bright excitonic peak is located at $\SI{1.47}{\electronvolt}$, corresponding to an exciton binding energy ($E_b$) of $\SI{0.72}{\electronvolt}$, computed as the difference between the QP direct gap and the electron-hole excitation energy. Additionally, a dark exciton lies nearly degenerate with this state within the thermal energy scale, $k_\text{B}T \approx \SI{25}{\milli\electronvolt}$. This dark state is found at $\SI{1.49}{\electronvolt}$, yielding an $E_b$ of $\SI{0.70}{\electronvolt}$. The second bright exciton appears at $\SI{1.82}{\electronvolt}$, with a corresponding $E_b$ of $\SI{0.37}{\electronvolt}$. Considering the same thermal energy window, an associated dark exciton is observed at $\SI{1.84}{\electronvolt}$, with an $E_b$ of $\SI{0.35}{\electronvolt}$. The highest-energy bound bright exciton emerges as a resonance at approximately $\SI{2.03}{\electronvolt}$, corresponding to a binding energy of $E_b = \SI{0.16}{\electronvolt}$, and is accompanied by several adjacent dark excitons within the thermal energy range. An analysis of its reciprocal-space wavefunction (not shown) reveals a complex structure with multiple nodal regions, suggesting a significant mixing of hydrogenic-like states. This nodal pattern indicates that the excitation cannot be described as a simple low-order hydrogenic state, but rather as a higher-order exciton resulting from the superposition of several excitonic configurations in reciprocal space.
\begin{figure*}[th!]
    \centering
    \includegraphics[width=\linewidth]{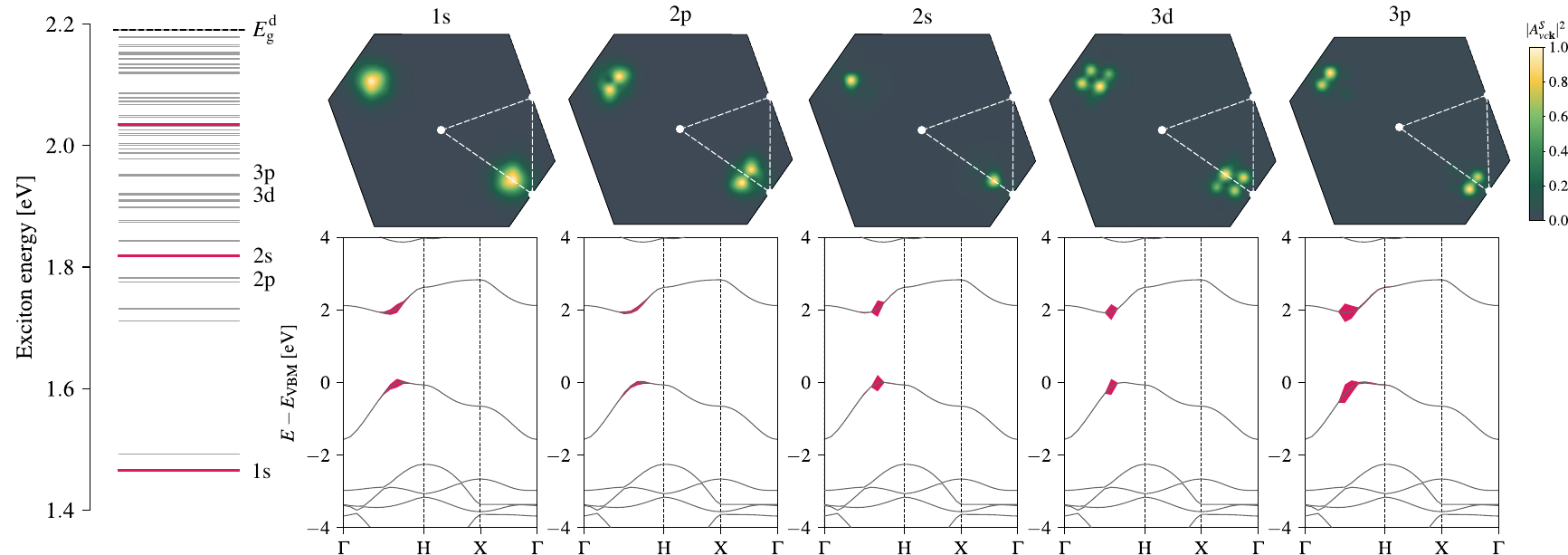}
    \caption{(Left) Energy spectrum of bound excitons in single-layer SGY. The GW direct band gap, corresponding to the continuum edge, is indicated by the dashed line at $\SI{2.19}{\electronvolt}$. Red (grey) lines denote bright (dark) excitons.
(Right, top) Exciton probability density in reciprocal space, given by the squared modulus of the exciton coefficients $\abs{A^{S}_{vc\mathbf{k}} }^{2}$, for the first few excitonic states of the hydrogenic series in SGY monolayer.
(Right, bottom) Quasiparticle GW band structure with the corresponding exciton weights in reciprocal space.}
    \label{fig:exciton_wf_reciprocal-space}
\end{figure*}

Strong Coulomb interactions lead to large spin-exchange energies that separate singlet and triplet excitons. Although spin-triplet (total spin $S=1$) excitons are typically dark in the single-photon optical absorption spectrum due to optical selection rules, the superfluidity of an exciton condensate can be probed through spin-transport experiments \cite{Jiang_PRL-2020}. Furthermore, spin mixing arising from weak spin–orbit coupling or hyperfine interaction can promote a transition from the singlet state to the optically dark triplet state, a process referred to as intersystem crossing \cite{Eizner_SA-2019}. In contrast to spin-singlet ($S=0$) excitons, triplet excitons carry spin when they flow, making the singlet–triplet energy splitting an important quantity for understanding exciton superfluidity. For the S-GY monolayer, we find that the lowest-energy spin-triplet exciton appears as a resonance at $\SI{1.35}{\electronvolt}$, corresponding to a singlet–triplet exchange splitting of approximately $\SI{120}{\milli\electronvolt}$. This value is comparable to the robust splitting of ${\sim}\SI{150}{\milli\electronvolt}$ reported for pristine GY \cite{Huang_PRB-2013}, and is about an order of magnitude larger than those typically observed in conventional semiconductors and carbon nanotubes \cite{Perebeinos_NL-2005}.

To gain further insight into the exciton spectrum, we plot the $\vb{k}$-space distribution of the squared amplitude of the bound excitonic states that compose the exciton wave functions in the Brillouin zone, allowing us to identify them within a Rydberg-like series. The results, together with the exciton energy spectrum, are shown in Fig.~\ref{fig:exciton_wf_reciprocal-space}. As in other two-dimensional semiconductors, the exciton energy levels do not strictly follow the Rydberg series predicted by the 2D hydrogenic model due to spatially nonuniform screening \cite{Antonius_NL-2018}. Nevertheless, from the nodal structure of the wave functions, we can identify a hydrogen-like sequence for the lowest-energy states with well-defined angular momentum character. This exciton series arises from band-to-band optical transitions near the lowest direct band gap along the $\Gamma$–H path, as evidenced by the quasiparticle GW band structure with the projected exciton weights. As shown by the reciprocal-space wave functions, the first two bright excitons correspond to $s$-like states, exhibiting an almost spherical symmetry. In between them, there are dark excitons with angular-momentum characters typical of $p$-like exciton states. These lowest-energy exciton states agree very well with the exciton spectrum of pristine GY \cite{Huang_PRB-2013}, although we can identify a larger number of dark excitons in S-GY. The higher-order excitons in the Rydberg series, $3d$ and $3p$, are both dark excitons.

\subsection{Exciton radiative lifetime}

Estimating exciton radiative lifetimes is fundamental for evaluating the optoelectronic potential of the S-GY monolayer and also represents a key requirement for achieving excitonic BEC. As a matter of fact, a long exciton lifetime favors the formation of thermodynamically equilibrated condensate states \cite{Mysyrowicz_PRL-1979}. However, this is often difficult to realize because exciton radiative recombination is frequently highly efficient, leading to very short lifetimes. The exciton radiative lifetimes at finite temperatures are estimated employing the model proposed by Chen \textit{et al.}~\cite{Chen_PRB-2019}, which has been suitable to achieve reliable lifetime predictions for many relevant 2D semiconductors \cite{Palummo_NL-2015,Villegas_PCCP-2024,Marinho_PRB-2025}. 

In this formalism, the radiative decay $\gamma_S(\vb{0})$, of an exciton state $S$ with  center-of-mass momentum $\vb{Q}=\vb{0}$, is described employing Fermi's golden rule as
\begin{equation}
    \gamma_S(\vb{0}) = \frac{1}{\tau_S(\vb{0})} = \frac{4\pi e^2}{\hbar^2c}\frac{\mu_S^2}{A_\text{uc}}\Omega_S(\vb{0})\,,\label{Eq:gamma_S}
\end{equation}
which corresponds to the inverse of the zero-temperature radiative lifetime $\tau_S(\vb{0})$. Here, \noindent $\mu_\text{S} = \sum_{vc\vb{k}} A^\text{S}_{vc\vb{k}}\hat{\vb{e}}\cdot\mel{c\vb{k}}{\vb{r}}{v\vb{k}}$ is the excitonic transition dipole matrix element, where $\hat{\vb{e}}$ and $\vb{r}$ are the polarization vector and the position operator, respectively, and $A^S_{vc\vb{k}}$ is the expansion coefficient of the $S$-th exciton eigenstate written as a superposition of electron and hole states $\ket{S\vb{Q}} = \sum_{vc\vb{k}}A^{S\vb{Q}}_{vc\vb{k}}a^{\dagger}_{c\vb{k}+\vb{Q}}a_{v\vb{k}}\ket{0}_{eh}$. $A_\text{uc}$ is the area of the unit cell, and $\Omega_S$ is the corresponding exciton eigenenergy.  Thus, the lifetime of an exciton in the $S$ state at temperature $T$ can be calculated by
\begin{equation}
    \tau_S(T) = \frac{3}{4}\tau_S(\vb{0})\left(\frac{2M_Sc^2}{\Omega_S(\vb{0})^2}\right)k_\text{B}T\,, \label{Eq:tau_S}
\end{equation}
where $M_S = m^*_e + m^*_h$ is the exciton effective mass. Afterwards, the effective radiative lifetime $\expval{\tau_\text{eff}(T)}$ is obtained by further averaging the rates from Eq.\,\eqref{Eq:tau_S} over the lowest-energy bright and dark excitons:
\begin{equation}
\expval{\tau_\text{eff}(T)} = \frac{\sum_S e^{-\Omega_S(\vb{0})/k_\text{B}T}}{\sum_S \left[\tau_S(T)\right]^{-1}e^{-\Omega_S(\vb{0})/k_\text{B}T}}\,.\label{Eq:tau_eff}
\end{equation}
The $\expval{\tau_\text{eff}(T)}$ represents the observable radiative recombination time of the exciton ensemble at temperature $T$, with excellent agreement with experimental measurements \cite{Palummo_NL-2015}. 

\begin{table*}[t!]
\caption{Exciton binding energies and radiative lifetimes for the excitonic states of the hydrogenic series in the SGY monolayer. The exciton radiative lifetimes are calculated at $\SI{0}{\kelvin}$ ($\tau^{0}$), at a low temperature of $\SI{4}{\kelvin}$ ($\tau^{\text{LT}}$), and at room temperature $\SI{300}{\kelvin}$ ($\tau^{\text{RT}}$).} The values in parentheses correspond to the radiative lifetimes of the dark exciton contributing to the 1s state.
\label{tab:excitons}
\begin{tabular}{lcccc}
\hline\hline
      & Exciton binding energy [eV] & $\tau^0$ [ps] & $\tau^\text{LT}$ [ns] & $\tau^\text{RT}$ [ns]\\
\hline
1s & $0.72$ & $0.12$ $(4.0)$ & $0.010$ $(0.29)$  & $0.66$ $(22)$\\
2p & $0.41$ & $21$ & $1.1$  & $151$ \\
2s & $0.37$ & $0.68$ & $0.033$  & $2.5$  \\
3d & $0.27$ & $53$ & $2.3$  & $334$  \\
3p & $0.24$ & $2.8$ & $0.12$ & $8.8$  \\\hline\hline
\end{tabular}
\end{table*}
\begin{figure}[t!]
    \centering
    \includegraphics[width=.6\linewidth]{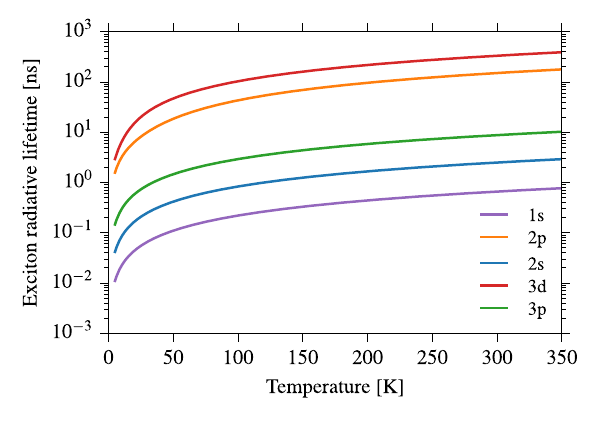}
    \caption{Temperature dependence of the exciton radiative lifetime for excitonic states in the Rydberg series for the S-GY monolayer.}
    \label{fig:radiative-lifetime_T}
\end{figure}
For the excitons in the Rydberg series identified in the S-GY monolayer, we compute the radiative lifetimes using Eq.\,\eqref{Eq:gamma_S} and Eq.\,\eqref{Eq:tau_S} at $T=\SI{0}{\kelvin}$, at low temperature ($\text{LT}=\SI{4}{\kelvin}$), and at room temperature ($\text{RT}=\SI{300}{\kelvin}$). The results, together with their corresponding binding energies, are reported in Table~\ref{tab:excitons}. At $\SI{0}{\kelvin}$, the estimated radiative lifetimes are on the order of $\SI{0.1}{\pico\second}-\SI{50}{\pico\second}$. Meanwhile, at low temperature ($\SI{4}{\kelvin}$) and room temperature ($\SI{300}{\kelvin}$), the radiative lifetimes in S-GY monolayers increase to the nanosecond scale, yielding values of ${\sim}\SI{0.01}{\nano\second}-\SI{300}{\nano\second}$. As observed, the radiative lifetimes of the dark excitons $2p$, $3d$, and $3p$ tend to be substantially longer than those of bright excitons. This is expected since the oscillator strength is proportional to the squared modulus of the exciton transition dipole moment $|\mu_S|^2$ \cite{Geiregat_LSA-2021}. Thus, from Eq.\,\eqref{Eq:gamma_S} and Eq.\,\eqref{Eq:tau_S}, the radiative lifetime is inversely proportional to the oscillator strength. Considering that the most intense resonance in the optical spectrum has an adjacent dark exciton, we also estimate the effective radiative lifetime by including the contribution of this dark state. This results in an increase of one to two orders of magnitude in the radiative lifetime. We further plot the exciton radiative lifetimes as a function of temperature, as shown in Fig.~\ref{fig:radiative-lifetime_T}. The radiative lifetimes increase monotonically with temperature for all excitonic states, reflecting the thermal occupation of excitons outside the light cone, which reduces the radiative recombination probability. In addition, dark excitons (2p, 3d, and 3p) exhibit consistently longer lifetimes than the bright 1s and 2s states, reaching the nanosecond-to-microsecond range at room temperature due to their suppressed optical transition dipole moments.

The exciton radiative lifetimes computed for the S-GY monolayer can be compared with those reported for other relevant low-dimensional semiconductors. In monolayer transition-metal dichalcogenides (TMDs), first-principles calculations and experiments find ultrafast radiative recombination for bright excitons at low temperature (tens of picoseconds), increasing to the nanosecond scale at room temperature due to thermal occupation of states outside the light cone~\cite{Palummo_NL-2015,Robert_PRB-2016,Chen_PRB-2019}. These trends are consistently verified in S-GY, where the bright $1s$ and $2s$ states present sub-picosecond intrinsic lifetimes at $T=\SI{0}{\kelvin}$ and nanosecond or longer effective lifetimes at $T=\SI{300}{\kelvin}$. In black phosphorus-based heterostructures, bright interlayer excitons are predicted to promote a reasonable balance between oscillator strength and radiative lifetime, although the radiative lifetimes lie in the picosecond regime.~\cite{Chen_2DM-2018}. For carbon nanotubes, a detailed first-principles analysis demonstrated that the intrinsic radiative lifetime of zero-momentum bright excitons is of the order of ${\sim}\SI{10}{\pico\second}$, while thermal population effects and the presence of low-lying dark excitons enhance the effective radiative lifetime to the ${\sim} \SI{10}{\nano\second}$ range at room temperature~\cite{Spataru_PRL-2005}. In addition, such long exciton radiative lifetimes have been predicted in monolayer graphitic carbon nitride (g-\ce{C3N4}) nanosheets, reaching values of the order of hundreds of nanoseconds at room temperature due to the presence of low-energy dark excitons~\cite{Fiorentin_ACSANM-2021}. In this scenario, our BSE-based radiative lifetimes for the excitons in the Rydberg series of the S-GY monolayer are analogous to those reported for other relevant 2D semiconductors, emphasizing the importance of dark excitons in yielding longer radiative lifetimes, while coexisting with bright excitons that provide a favorable balance between oscillator strength and lifetime.

To further characterize exciton decay in the S-GY monolayer under realistic experimental conditions, we compute the effective radiative lifetime $\expval{\tau_{\mathrm{eff}}}$ as defined in Eq.\,\eqref{Eq:tau_eff}. In this approach, the recombination rates of the excitonic ensemble are thermally averaged according to a Boltzmann distribution, thus accounting for the thermal population of both bright and dark excitons, as well as excitons lying outside the light cone. At room temperature ($T=\SI{300}{\kelvin}$), including all bound excitons below the continuum edge, we obtain $\expval{\tau_{\mathrm{eff}}}=\SI{0.89}{\nano\second}$ for S-GY. Such an effective radiative lifetime is comparable to the experimental exciton lifetimes reported for \ce{MoS2} ($\SI{0.85}{\nano\second}$) and \ce{MoSe2} ($\SI{0.90}{\nano\second}$) monolayers \cite{Shi_ACSN-2013,Kumar_PRB-2014}. This value indicates that radiative recombination in S-GY remains efficient under ambient conditions, while still being influenced by the presence of nearby dark excitonic states, which act as long-lived reservoirs and contribute to extending the overall exciton decay time. Therefore, the computed $\expval{\tau_{\mathrm{eff}}}$ reinforces that S-GY combines a favorable balance between oscillator strength and radiative lifetime.

\subsection{Phase diagram of electron-hole pairs}
Large binding energies, together with long lifetimes, are key features for the emergence of macroscopic quantum phenomena involving excitons. These characteristics ensure that excitons maintain their quasiparticle nature during coherent condensation. The binding energy $E_b = \SI{0.72}{\electronvolt}$ of the 1s exciton in S-GY suggests that it can remain stable at relatively high temperatures. Empirically, the thermal energy associated with exciton dissociation is about $\SI{10}{\percent}$ of its binding energy, i.e., $k_\text{B}T_d \approx 0.1E_b$ \cite{Keldysh_CP-1986}. In this sense, the exciton dissociation temperature $T_d$ corresponds to the temperature above which an exciton is no longer stable as a bound electron-hole pair. Based on this estimate, the computed $T_d$ is approximately $\SI{840}{\kelvin}$ for the 1s exciton in S-GY monolayer, significantly above room temperature. We stress that this dissociation temperature $T_d$ should be regarded as an upper bound, since the presence of a substrate is expected to reduce the exciton binding energy $E_b$ \cite{Rustagi_NL-2018}. Accordingly, $T_d$ in supported samples is expected to be smaller than that of a free-standing monolayer.

Another crucial parameter governing the quantum phases of electron-hole pairs is the exciton Bohr radius $a_S$, which characterizes the spatial extent of the exciton and determines the density range over which a dilute exciton gas can behave as a stable ensemble of composite bosons before significant wavefunction overlap leads to dissociation \cite{Combescot_RPP-2017}.

The 1s exciton in S-GY is modeled within the effective mass approximation by employing the Wannier--Mott two-particle Schr\"odinger equation in its 2D hydrogenic form \cite{Rodin_PRB-2014}:
\begin{equation}
    \left[-\frac{\hbar^2}{2\mu_{eh}}\laplacian_\text{2D}+\Phi_\text{2D}(r)\right]\psi_S(\vb{r}) = -E_b \psi_S(\vb{r})\,,\label{eq:Schrodinger_Eq_ex}
\end{equation}
where $r = \abs{\vb{r}_e - \vb{r}_h}$ is the separation between the electron and hole, and $\mu_{eh} = (m^*_e m^*_h)/(m^*_e + m^*_h)$ is the reduced mass of the electron-hole pair. For free-standing monolayers, the screening effect is captured by the Rytova-Keldysh potential $\Phi^\text{RK}_\text{2D}(r)$ \cite{Wang_RMP-2018}:
\begin{equation}
\Phi^\text{RK}_\text{2D}(r) = -\frac{e^2}{4\pi\epsilon_0}\frac{\pi}{2\kappa r_0}\left[\vb{H}_0\left(\frac{r}{r_0}\right) - Y_0\left(\frac{r}{r_0}\right)\right]\,,\label{eq:RK-potential}    
\end{equation}
where $\kappa$ is the average relative dielectric constant of the environment, $\vb{H}_0(x)$ and $Y_0(x)$ are the Struve and Bessel functions of the second kind, respectively, and $r_0$ is the screening length. For an isolated S-GY monolayer, $\kappa = 1$ and $r_0 = 2\pi \alpha_\text{2D}/\kappa$ \cite{Rodin_PRB-2014}, where $\alpha_\text{2D}$ is the two-dimensional polarizability. Importantly, $\kappa$ is the average dielectric constant of the surrounding medium and is set to unity for a freestanding monolayer. We note that this choice corresponds to an idealized unscreened limit, and environmental screening from substrates or encapsulation is expected to reduce the exciton binding energy and quantitatively modify the excitonic phase boundaries. For a vacuum length of $L=\SI{15}{\angstrom}$, $\alpha_\text{2D}$ is estimated as $\SI{3.48}{\angstrom}$. As verified, this value is already converged, since $\alpha_\text{2D}$ changes by only about $\SI{0.6}{\percent}$ when $L$ is increased from $\SI{10}{\angstrom}$ to $\SI{25}{\angstrom}$. This results in a $r_0$ of $\SI{21.86}{\angstrom}$. At this point, we can fully specify the Rytova-Keldysh potential and solve the effective Schrödinger equation [Eq.~\eqref{eq:Schrodinger_Eq_ex}] by employing the variational method \cite{Quintela_PSS-2022,Yue_PRB-2025}. To this end, the 2D Slater-type ansatz for the 1s exciton variational wavefunction reads
\begin{equation}
    \psi_S(r) = \sqrt{\frac{2}{\pi \lambda^2}}e^{-r/\lambda}\,.\label{eq:ansatz}
\end{equation}
where $\lambda$ is the variational parameter. After minimizing the energy with respect to the variational parameter, we obtain $\lambda = \SI{10.24}{\angstrom}$. For the adopted variational ansatz, the expectation value of the relative coordinate satisfies 
$a_S = \mel{\psi_S}{\hat{r}}{\psi_S} = \lambda$, such that the variational parameter directly represents the average electron–hole separation \cite{Yue_PRB-2025}, and therefore the effective exciton Bohr radius is $a_S = \SI{10.24}{\angstrom}$.  The obtained $a_S$ for the 1s exciton in S-GY is comparable to typical values reported for monolayer TMDCs (${\sim}\SI{1}{\nano\meter}$) \cite{Wang_RMP-2018}. For validation purposes, the corresponding $E_b$ is $\SI{0.73}{\electronvolt}$, in excellent agreement with the value of $\SI{0.72}{\electronvolt}$ obtained within the \textit{ab initio} GW$+$BSE framework.

\begin{figure}[t!]
    \centering
    \includegraphics[width=.6\linewidth]{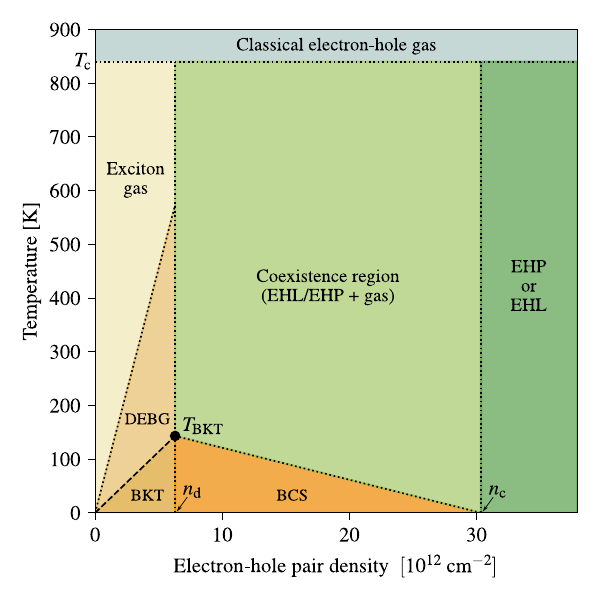}
    \caption{Temperature--density phase diagram for coherent excitonic phases of the 1s exciton in S-GY monolayer. In the dilute regime ($n<n_d$), the system exhibits a classical exciton gas at high temperature, a quantum-degenerate exciton Bose gas (DEBG) below the quantum-degeneracy line, and a BKT superfluid phase for $T<T_\text{BKT}$. For $n>n_d$, exciton overlap leads to a regime where electron-hole liquid (EHL) or electron-hole plasma (EHP) coexists with exciton gas, and a BCS-like paired phase at low temperature. Beyond the Mott critical density $n_c$, the system crosses into an electron--hole plasma/liquid (EHP/EHL) regime.}
    \label{fig:phase-diagram-excitons}
\end{figure}

In addition to the critical temperature $T_\text{c}$, the Mott critical density $n_\text{c}$ is a fundamental parameter for determining the stability of the excitonic phase, given that above this critical density, Coulomb screening and Pauli blocking suppress bound excitons and the system evolves into an electron-hole liquid or plasma (EHL/EHP) regime. For 2D systems, the Mott critical density can be estimated from the overlap criterion as \cite{Deng_RMP-2010,Cudazzo_PRL-2010}
\begin{equation}
    n_\text{c} = \frac{1}{\pi a_S^2}\,.
\end{equation}
 Hence, we obtain $n_\text{c} = \SI{30.4e12}{\centi\meter^{-2}}$ for the 1s exciton in S-GY.

To further characterize the low-density regime, we also calculate the dilute density $n_\text{d}$ from the Fisher-Hohenberg criterion~\cite{Fisher_Hohenberg_PRB-1988}, defined by
\begin{equation}
    \ln\!\left[\ln\!\left(\frac{1}{n_\text{d} a_S^2}\right)\right] = 1\,.
\end{equation}

This condition determines the density above which the exciton gas can no longer be treated as a weakly interacting dilute two-dimensional Bose system. Unlike $n_\text{c}$, the density $n_\text{d}$ does not correspond to exciton dissociation, but rather to the onset of strong interaction effects within the excitonic phase. We estimate $n_\text{d} = \SI{6.3e12}{\centi\meter^{-2}}$. For densities between $n_\text{d}$ and $n_\text{c}$, the system progressively crosses over into a coexistence region where droplets of EHL, EHP and the exciton gas are simultaneously present in thermodynamic equilibrium \cite{Rustagi_NL-2018,Klingshirn_SO-book_2012}, and at sufficiently low temperatures, it may further evolve into the correlated Bardeen-Cooper-Schrieffer (BCS)-like phase characterized by electron–hole Cooper pairs.

Based on the characteristic scales $T_\text{c}$, $n_\text{c}$, and $n_\text{d}$, we construct the phase diagram of the coherent excitonic phases for the 1s exciton in the S-GY monolayer, as presented in Fig.~\ref{fig:phase-diagram-excitons}. 

The onset of the degenerate exciton Bose gas (DEBG) regime is determined from the quantum degeneracy condition of a 2D Bose gas \cite{Fogler_NC-2014},
\begin{equation}
    n_{\mathrm{DEBG}}(T) = \frac{M_Sk_B T }{2\pi \hbar^2}\,.
\end{equation}
This relation defines the density-temperature boundary separating the classical exciton gas from the quantum-degenerate regime. Quantum degeneracy occurs when $n > n_{\mathrm{DEBG}}(T)$, i.e., when the thermal de Broglie wavelength becomes comparable to the interparticle spacing. The corresponding degeneracy temperature $T_\text{d}$ follows from the condition $n_\text{d} = n_{\mathrm{DEBG}}(T_\text{d})$, yielding $T_\text{d} \approx \SI{573}{\kelvin}$.

In two dimensions, Bose–Einstein condensation does not occur at finite temperature in either ideal or interacting systems \cite{Fisher_Hohenberg_PRB-1988}. Nevertheless, as the temperature decreases, a phase transition to a superfluid state can take place via the Berezinskii–Kosterlitz–Thouless (BKT) mechanism \cite{berezinskii_JETP-1971,Kosterlitz_JPSSP-1972}. In this phase, superfluidity emerges from the binding of vortex–antivortex pairs, leading to quasi-long-range order and dissipationless flow of composite bosons \cite{Ryzhov_PU-2017}. The BKT transition density as a function of temperature is given by
\begin{equation}
    n_\text{BKT}(T) = \frac{2 M_S k_\text{B} T}{\pi \hbar^2}\,,
\end{equation}
By imposing $n_\text{BKT} = n_\text{d}$, we obtain a maximum superfluid transition temperature of $T_\text{BKT} \approx \SI{143}{\kelvin}$ for the 1s exciton in S-GY monolayer.

Such a remarkable $T_\text{BKT}$ indicates the potential for excitonic superfluidity at high temperatures, being comparable to the temperatures reported for exciton condensation in 2D atomic double layers and consistent with theoretical predictions of superfluidity of indirect excitons in van der Waals heterostructures \cite{Guo_ScienceAdv-2022, Fogler_NC-2014}. The predicted superfluid transition temperature for S-GY is quite similar to the $\SI{145}{\kelvin}$ recently reported for monolayer \ce{TiS3} \cite{Wang_JPCL-2021}, thus placing S-GY among the promising candidates for high-temperature exciton superfluidity. Interestingly, a recent theoretical study on holey graphyne \cite{Yue_PRB-2025}, a member of the GBM family closely related to S-GY, predicted high-temperature superfluidity for dark excitons, with a predicted $T_\text{BKT}$ of about $\SI{126}{\kelvin}$, highlighting that GY-based carbon allotropes may host robust excitonic many-body phases at elevated temperatures.

Nevertheless, the phase diagram of electron–hole pairs shown in Fig.~\ref{fig:phase-diagram-excitons} is obtained under the assumption of a freestanding monolayer. We emphasize that, under realistic experimental conditions, the presence of a substrate or dielectric environment enhances screening, thereby reducing the exciton binding energy and, consequently, the estimated BKT transition temperature. Therefore, the reported transition temperatures should be regarded as upper bounds. In fact, the overall trends and the possibility of accessing excitonic collective phases are expected to remain robust, particularly in weakly screened or suspended configurations.

The expected excitonic spectrum and the phase diagram of electron–hole pairs provide clear experimental signatures that can be probed using standard spectroscopic techniques. For example, the large binding energies and non-hydrogenic Rydberg series can be directly measured via optical absorption or photoluminescence. In addition, the exciton energies depend on the dielectric environment, enabling their tuning through substrate engineering or dielectric encapsulation. Furthermore, the predicted BKT transition temperatures suggest that low-temperature transport or Coulomb drag experiments may reveal signatures of excitonic collective phases. These results provide clear guidance for future experimental investigations of excitonic effects in graphyne-based systems.

\section{\label{{sec:conclusions}} Conclusions}
We have carried out a comprehensive investigation of the quasiparticle electronic structure, optical response, and excitonic properties of monolayer S-GY using the many-body perturbation theory within the GW--BSE approach. Our results reveal that quasiparticle corrections significantly renormalize the band gap while preserving relatively light carrier effective masses. The excitonic spectrum is dominated by strongly bound excitons with large binding energies and clear deviations from the conventional hydrogenic Rydberg series, driven by reduced dimensionality and nonlocal screening effects. The presence of a nearly degenerate dark exciton state, together with radiative lifetimes in the nanosecond range at room temperature, highlights the potential for enhanced exciton stability in this system.

By combining GW--BSE calculations with an effective Rytova--Keldysh model, we established a consistent description of the excitonic behavior and its environmental dependence. Beyond single-exciton properties, we identified the conditions for collective excitonic phases. In particular, we determined a dilute density of $\SI{6.3e12}{\centi\meter^{-2}}$, below which the system behaves as a rarefied two-dimensional Bose gas. Within this regime, the estimated Berezinskii--Kosterlitz--Thouless transition temperature reaches a value as high as $\SI{143}{\kelvin}$, indicating that excitonic superfluidity could emerge at experimentally accessible temperatures under weak screening conditions. Such a remarkably high estimated transition temperature, together with the large binding energy and long exciton lifetimes, suggests that S-GY is a promising platform for exploring excitonic superfluidity at elevated temperatures, thereby expanding the landscape of GY-based materials toward experimentally relevant high-temperature excitonic regimes.

\begin{acknowledgement}
The authors thank the ``Centro Nacional de Processamento de Alto Desempenho em São Paulo'' (CENAPAD-SP, UNICAMP/FINEP – MCTI project) for support related to projects No.~634, 897, and 974. A.C.D acknowledges the financial support from Federal District Research Support Foundation (FAPDF, grants $00193-00001817/2023-43$ and $00193-00002073/2023-84$), the National Council for Scientific and Technological Development $-$ CNPq ($408144/2022-0$, $305174/2023-1$, $444431/2024-1$, and $444069/2024-0$). A.C.D and L.A.R.J acknowledge PDPG-FAPDF-CAPES Centro-Oeste $00193-00000867/2024-94$. L.A.R.J. acknowledges the financial support from FAP-DF grants 00193.00001808/2022-71 and $00193-00001857/2023-95$, FAPDF-PRONEM grant 00193.00001247/2021-20, and CNPq grants $350176/2022-1$, $167745/2023-9$, $444111/2024-7$. 
\end{acknowledgement}


\bibliography{references}

\end{document}